\documentclass[
 reprint,
 amsmath,amssymb,
 aps,
]{revtex4-2}

\usepackage{graphicx}
\usepackage{bm}

\usepackage[colorlinks=true, linkcolor=blue,anchorcolor=blue, citecolor=blue,urlcolor=blue]{hyperref}

\usepackage[normalem]{ulem}

\usepackage{xcolor}

\begin{document}
\preprint{APS/123-QED}

\title{Tunable Edelstein effect in intrinsic two-dimensional ferroelectric metal PtBi$_{2}$}
\author{Weiyi Pan$^{1}$}
\email{Weiyi.Pan@physik.uni-regensburg.de}
\author{Jaroslav Fabian$^{1,2}$}

\affiliation{$^{1}$Institute for Theoretical Physics, University of Regensburg, 93040 Regensburg, Germany\\
$^{2}$Halle-Berlin-Regensburg Cluster of Excellence CCE, Halle, Germany\\
}

\begin{abstract}
The Edelstein effect, which enables charge-to-spin conversion and is therefore highly promising for future spintronic devices, can be realized and non-volatilely manipulated in ferroelectric materials owing to their broken inversion symmetry and switchable polarization states. To date, most ferroelectric systems reported to exhibit the Edelstein effect are semiconductors, requiring extrinsic doping for functionality. In contrast, the Edelstein effect has rarely been reported in metallic ferroelectric systems, where doping is unnecessary. Using first-principles calculations, we predict that a pronounced  Edelstein effect can be realized in the recently proposed intrinsic two-dimensional ferroelectric metal PtBi$_{2}$ monolayer, where the sign of the Edelstein coefficient is coupled to the direction of ferroelectric polarization through the polarization-switching–induced reversal of spin textures, thereby enabling non-volatile control of charge–spin conversion. The Edelstein effect reaches a magnitude of $10^{11}~\hbar/(\textup{A} \cdot \textup{cm})$, which is sizable compared to previously reported ferroelectric systems. Microscopically, the Edelstein effect in a PtBi$_2$ monolayer originates from competing contributions of inner Rashba-like electron pockets and outer hole pockets with opposite signs; an upward shift of the Fermi level alters their balance and can reverse the sign of the Edelstein effect. Upon applying biaxial strain, the Fermi-surface electronic structure is strongly modified, resulting in a pronounced change of the Edelstein effect: a 2 \% compressive strain suppresses the Edelstein effect by about 50 \%. Our results not only identify a promising material platform for tunable charge–spin conversion but also provide new insights into the functional potential of metallic ferroelectric systems.

\end{abstract}
\maketitle
\section{Introduction}
The Edelstein effect (EE)\cite{EDE,Johansson_2024}, which refers to the conversion of an applied charge current into a nonequilibrium spin accumulation in the presence of spin orbital coupling (SOC), plays a central role in the development of spin–orbit-torque devices and spin-based logic architectures. Quantitatively, the EE can be described by a susceptibility tensor $\chi$, which links the induced spin polarization $\delta \textbf{S}$ to the applied electric field $\textbf{E}$\cite{Xu,Johansson_2024}: $\delta s_{i} = \chi_{ij} E_{j}$. Such EE, which typically originates from noncollinear spin textures in k-space generated by SOC, requires broken inversion symmetry and would experience a sign reversal upon the spatial inversion operation. Therefore, achieving a sizable EE in practical material systems hinges on simultaneously breaking inversion symmetry and ensuring the presence of strong SOC.

Previously, the EE has been investigated in a variety of inversion-asymmetric systems, including two-dimensional (2D) van der Waals heterostructures\cite{REE1,REE2,REE3,REE4,REE5}, metal interfaces\cite{REE6,REE7}, 2D electron gases at oxide interfaces\cite{REE8,REE9,REE10,REE11}, surfaces of topological insulators\cite{REE12,REE13,REE14}, and chiral nanowires\cite{REE15,REE16,REE17}. In these systems, the spin textures can be further tuned through voltage gating\cite{REE16} or twisting\cite{REE3,REE5}, thereby modulating the magnitude of the EE. 

Ferroelectric materials offer a promising platform for studying the EE\cite{FE1,FE2,FE3,FE4,FE5,FE6,FE7}. In ferroelectric systems, the absence of inversion symmetry naturally allows the noncollinear k-space spin textures and the EE to occur. Moreover, by switching the direction of electric polarization, both the helicity of spin textures and the sign of EE coefficient can be inverted, which provides opportunities for designing non-volatile and electrically controllable device functionalities. Previously, ferroelectric-controllable EE has been explored in several ferroelectric systems, such as GeTe\cite{FE1,FE7}, In$_{2}$Se$_{3}$\cite{FE3}, graphene/In$_{2}$Se$_{3}$ heterostructure\cite{MIS}, and CsGeX$_{3}$ (X = I, Br, Cl)\cite{FE6}. However, most observed ferroelectric materials are insulating, making it necessary to dope them into a conductive regime to realize the EE and thus increase the burden of the experiment. In contrast, emerging ferroelectric metals—where ferroelectricity coexists with metallicity—offer a new route toward achieving switchable EE without the need for doping\cite{FE8,FE9,FE10}. Nevertheless, intrinsic ferroelectric metals, particularly those in the 2D limit that are advantageous for both highly integrated electronic applications and higher tunability for potential functionalities, remain largely elusive. Therefore, more candidates of intrinsic 2D ferroelectric metals with effective EE are highly desired both theoretically and experimentally. 

\begin{figure*}[ht]
\includegraphics[scale = 0.28 ]{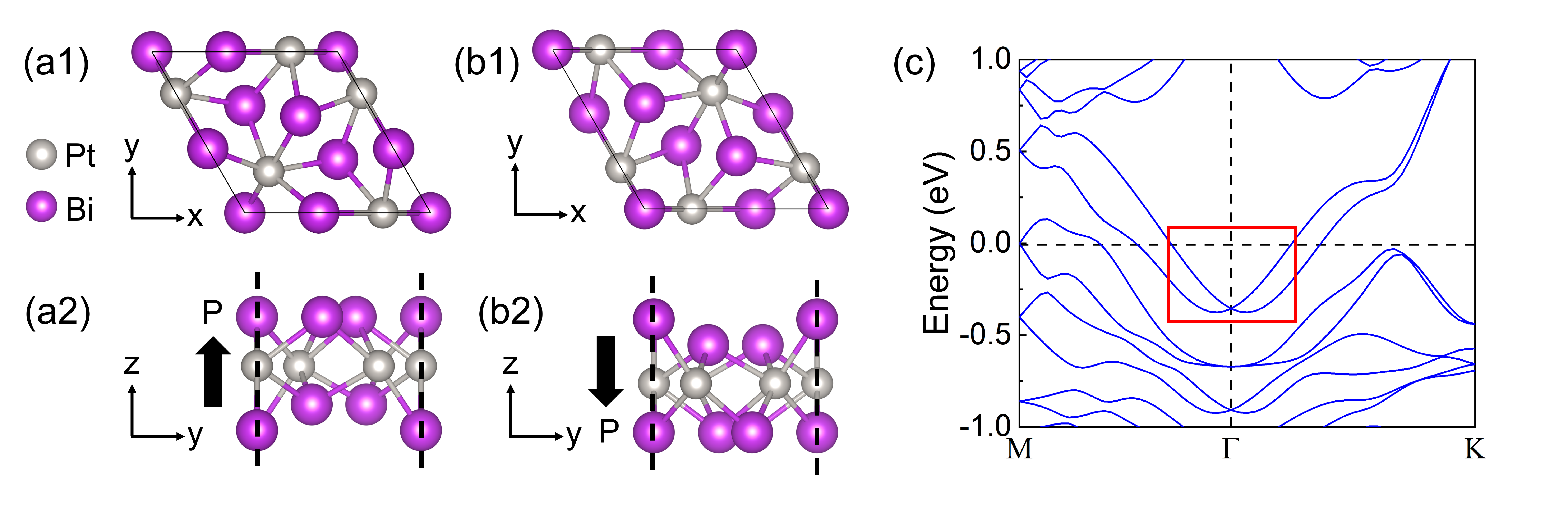}
\caption{\label{1} The atomic structure of monolayer PtBi$_{2}$ with the polarization (a) upwards (P$_{\uparrow}$ state) and (d) downwards (P$_{\downarrow}$ state). (a1) and (b1) show the top view while (a2) and (b2) show the side view. (c) The band structure of monolayer PtBi$_{2}$ with polarization upwards, while the band structure of the PtBi$_{2}$ with polarization downwards is identical. The Rashba-like bands forming electron pockets near the $\Gamma$ point are circled.  }
\end{figure*}

Recently, monolayer PtBi$_{2}$ has  been proposed as a rare example of an intrinsic 2D ferroelectric metal\cite{Wu} (see Fig. \ref{1}). It has been predicted that monolayer PtBi$_{2}$ can be exfoliated from experimentally accessible bulk trigonal PtBi$_{2}$\cite{PhysRevB.94.165119,PhysRevB.94.235140,PhysRevMaterials.4.124202,changdar2025topological} with moderate exfolation energy\cite{Wu}, and it also exhibits robust metallicity together with reversible ferroelectric polarization originated from distorted monolayer structure, featuring a moderate switching barrier and a considerable polarization magnitude\cite{Wu}. Moreover, the strong SOC inherent to PtBi$_{2}$ gives rise to a topologically nontrivial band structure\cite{Wu}, which in turn leads to an enhanced shift-current conductivity desirable for future photovoltaic applications\cite{PhysRevLett.116.237402}. This unique combination of intrinsic ferroelectricity, metallicity, and strong SOC in monolayer PtBi$_{2}$ makes it an excellent candidate for realizing intrinsic and switchable EE, which motivates us to conduct a detailed investigation of its EE performance, tunability, and the microscopic mechanisms underlying it.

In this work, we systematically investigate the tunable EE in monolayer PtBi$_{2}$. Based on first-principles calculations, we demonstrate that a considerable EE can intrinsically exist in this material. Such EE reaches the magnitude of $10^{11} \hbar/(\textup{A} \cdot \textup{cm})$, which represents a relatively large value among previously reported ferroelectric systems. Upon reversing the ferroelectric polarization of monolayer PtBi$_{2}$, the sign of the EE coefficient also switches. Microscopically, the EE arises from combined contributions of the inner Rashba-like electron pockets and the outer hole pockets on the Fermi surface, which may compete with one another. Furthermore, the EE in monolayer PtBi$_{2}$ can be additionally tuned by applying biaxial strain. Our findings not only provide new insights into the functional potential of intrinsic ferroelectric metals, but also broaden the range of candidate materials for charge–spin conversion in future spintronic devices.

\section{Methods}

Our first-principles calculations are performed using the Quantum Espresso package\cite{QE}. In our work, the Perdew–Burke–Ernzerhof (PBE)\cite{PBE} exchange–correlation functional and projector-augmented-wave (PAW)\cite{PAW} potentials are employed. Both structural relaxation and self-consistent calculations use a $9\times9\times1$ k-point mesh, with energy cutoffs of 500 Ry for the charge density and 60 Ry for the wavefunctions. Structural relaxations are carried out using scalar-relativistic pseudopotentials and the quasi-Newton algorithm until all components of the forces are smaller than 3×10$^{-4}$ Ry/$a_{0}$, where $a_{0}$ is the Bohr radius. Self-consistent calculations are then performed with fully relativistic pseudopotentials to account for spin–orbit coupling (SOC). A vacuum spacing of 28 \AA \ is applied in all simulations to eliminate interactions between periodic images in the slab geometry.

After the DFT calculations, we construct the Wannier Hamiltonian of our system using the Wannier90\cite{w901} code. For the wannierization procedure, the $d$-orbitals of Pt and the $p$-orbitals of Bi are selected as the projection basis. Based on the resulting Wannier Hamiltonian, we employ the Wanniertools\cite{wtools} package to compute the Fermi surface and the corresponding spin textures. In addition, we use the Linres\cite{linears} code to evaluate the Edelstein effect (EE) via the Kubo formula. Specifically, we consider the linear response of spin polarization to an external electric field, expressed as $\delta \mathbf{S}=\chi\mathbf{E}$, where $\delta \mathbf{S}$ is the induced spin polarization, $\mathbf{E}$ is the electric field, and $\chi$ is the response tensor in nonmagnetic systems, which can be written as\cite{Kubo1,Kubo2,PhysRevB.110.214419}:
\begin{equation}
    \chi_{ij} = -\frac{e\hbar}{\pi  N} \sum_{\mathbf{k},m,n} \frac{\Gamma^{2} \textup{Re}(\langle n\mathbf{k} |\hat{S}_{i} | m\mathbf{k} \rangle \langle m\mathbf{k} | \hat{v}_{j} | n\mathbf{k} \rangle  )  }{[(E_{f}-\epsilon_{n\mathbf{k}})^{2}+ \Gamma^{2}][(E_{f}-\epsilon_{m\mathbf{k}})^{2}+ \Gamma^{2}]}
\end{equation}

Here, $e$ is the elementary charge; $n$ and $m$ denote band indices; $\textbf{k}$ is the Bloch vector; $E_{F}$ is the Fermi energy; $\hat{v}$ is the velocity operator, $\hat{S}$ is the spin operator; $\epsilon_{n\textbf{k}}$ is the eigenvalue; $V$ is the volume of unit cell; $N$ is the total number of k points used to sample the Brillouin zone; and $\Gamma$ is the disorder parameter, which is related to the relaxation time $\tau$ through $\tau = \hbar/2\Gamma$. It is worth noting that the EE originates mainly from the electronic states on the Fermi surface\cite{PhysRevLett.128.166601}. This becomes evident when $\Gamma$ is small, since in this case the above expression can be approximated as\cite{Kubo1}:
\begin{equation}
    \chi_{ij} = -\frac{e\hbar}{2 \Gamma  N} \sum_{\textbf{k},n}     \delta(\epsilon_{nk}- E_{F}) 
 \langle n\textbf{k} | \hat{S}_{i} | n\textbf{k} \rangle \langle n\textbf{k} |\hat{v}_{j} | n\textbf{k} \rangle  
\end{equation}

where the $\delta(\epsilon_{n\textbf{k}}- E_{F})$ part explicitly indicates that states at the Fermi level will dominate the EE. Here we set $\Gamma = 10$ meV, which roughly corresponds to a typical sub-100 fs scattering rate (broadening) while remaining much smaller than interband separations.
Finally, a $400\times400\times1$ k-point mesh is used to obtain the converged EE coefficient. 


\section{RESULTS AND DISCUSSION}

The monolayer PtBi$_{2}$ crystallizes in space group No. 157 ($C_{3v}$ point group). It consists of a Pt layer sandwiched between two Bi layers, forming a distorted structure with nine atoms in the unit cell, which is shown in Fig. \ref{1}. The in-plane lattice constant is calculated to be 6.56 \AA, which is consistent with previous studies \cite{Wu,PhysRevB.94.165119}. Such monolayer PtBi$_{2}$ is reported to have two ferroelectric ground state with opposite electric polarizations, namely, P$_{\uparrow}$ and P$_{\downarrow}$ states (see Figs. \ref{1}(a) and \ref{1}(b)), which are connected by the inversion operation $\mathcal{P}$ and are electrically switchable\cite{Wu}. In the P$_{\uparrow}$ (P$_{\downarrow}$) configuration, the top (bottom) Bi layer lies in a single plane, whereas the Bi atoms in the bottom (top) layer do not lie in the same plane. Such a structural arrangement breaks both inversion symmetry $\mathcal{P}$ and mirror symmetry $\mathcal{M}_{z}$, which allows the existence of perpendicular electric polarization\cite{Wu}.
 

To gain further insights, we evaluate the electronic band structure of monolayer PtBi$_{2}$, as shown in Fig. \ref{1}(c). Several energy bands cross the Fermi level, indicating pronounced metallic behavior. Notably, two bands intersect the Fermi level in the vicinity of the $\Gamma$ point, forming two electron pockets. These bands are degenerate at the $\Gamma$ point but split upon moving away from it, exhibiting a characteristic Rashba-type band splitting that could potentially be observed experimentally via angle-resolved photoemission spectroscopy (ARPES)\cite{AREPS,ARPES2}. Quantitatively, the Rashba effect can be described by the Rashba parameter $\alpha_R =2E_R/k_R$, where $E_R$ and $k_R$ denote the Rashba splitting energy and momentum offset, respectively\cite{AREPS}. The calculated Rashba parameter is $\alpha_R=1.10 eV \cdot $\AA, which is considerable compared with several representative 2D Rashba systems, such as metal surface states\cite{PhysRevLett.77.3419,PhysRevLett.93.046403}, 2D electron gases \cite{PhysRevLett.107.096802,REE9}, and 2D heterostructures\cite{yi2022crossover,AREPS}. Note that such Rashba splitting is usually expected to have significant contribution to EE\cite{Rashba,Johansson_2024}. In addition to the two electron pockets near the $\Gamma$ point, another band crosses the Fermi level near the $M$ point, giving rise to hole pockets on the Fermi surface.

 \begin{figure}[ht]
\includegraphics[scale = 0.45 ]{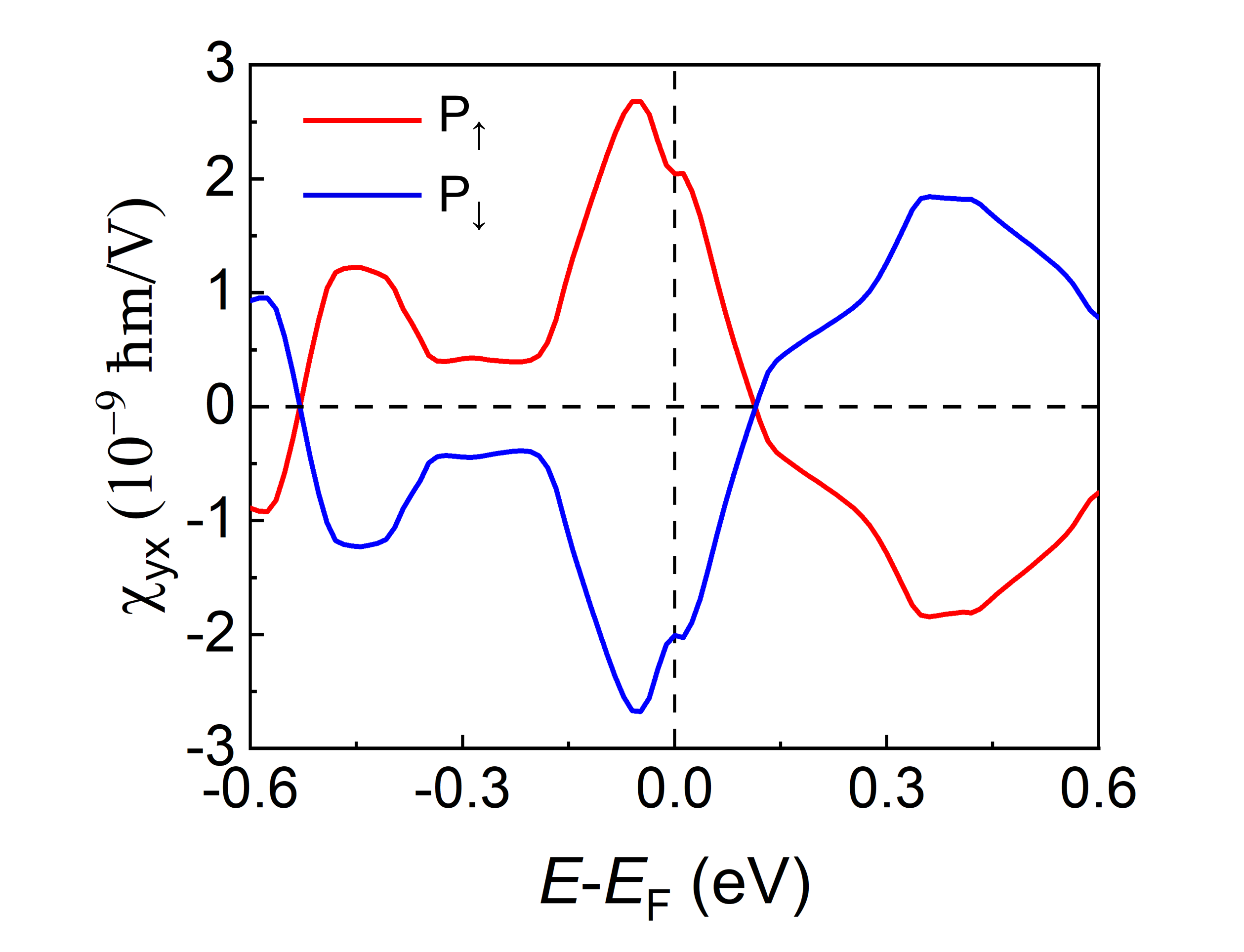}
\caption{\label{2} Calculated EE coefficient in monolayer PtBi$_{2}$. The curves corresponding to opposite polarization states P$_{\uparrow}$ and P$_{\downarrow}$ are labelled in red and blue, respectively.}
\end{figure}

 \begin{figure*}[ht]
\includegraphics[scale = 0.38 ]{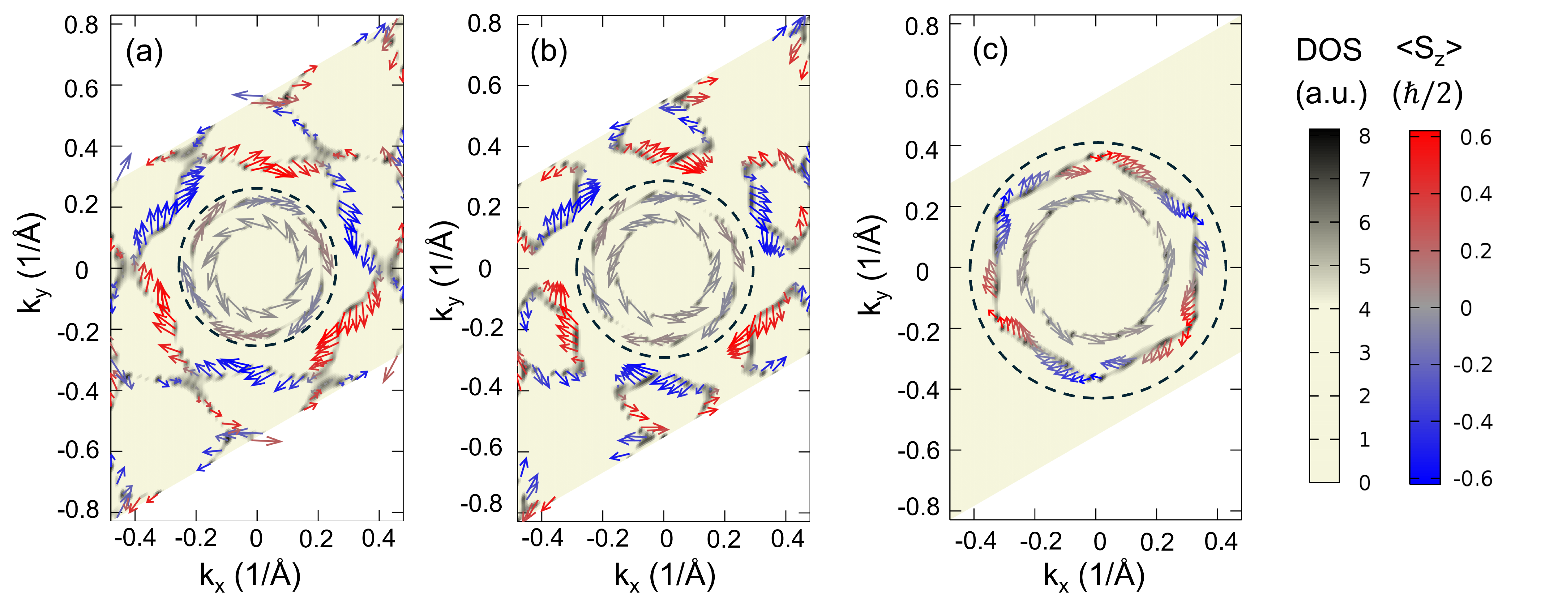}
\caption{\label{3} The Fermi surface states as well as the corresponding spin textures for P$_{\uparrow}$ state of monolayer PtBi$_{2}$ at (a) $E_{F}$ = -0.05 eV, (b) $E_{F}$ = 0 eV, and (c) $E_{F}$ = 0.2 eV, respectively. The background color shows the spectral density of electronic states on the Fermi surface, expressed in arbitrary units. The arrows indicate the direction and magnitude of the in-plane spin expectation values for these Fermi surface states. The color of the arrow represents the expectation value of out-of-plane spin components. The states from inner electron pockets near the $\Gamma$ point are circled, while the states out of the circle are from outer hole pockets. }
\end{figure*}

Since monolayer PtBi$_{2}$ exhibits both inversion-symmetry breaking and strong SOC, the presence of the EE is naturally expected. Moreover, in ferroelectric systems, the sign of the EE coefficient can be reversed through ferroelectric switching, which may also occur in monolayer PtBi$_{2}$. Based on symmetry considerations, the space group No. 157 ($C_{3v}$ point group) allows only two nonzero components of the EE tensor, $\chi_{yx}=-\chi_{xy}$, to exist in monolayer PtBi$_{2}$. By computing the $\chi_{yx}$ component of the EE coefficient as a function of chemical potential and polarization direction (see Fig. \ref{2}), one finds that reversing the polarization indeed switches the sign of the EE coefficient. Note that ferroelectric-switchable EE has been reported previously in ferroelectric semiconductors\cite{FE3,FE6}. In contrast, for metallic ferroelectrics, a finite EE can exist without the need for additional doping. Although a previous theoretical study has discussed this effect in 1T$^{'}$-WTe$_2$ \cite{FE4}, more reports of EE in ferroelectric metals remain scarce and are therefore highly desirable. For the P$_{\uparrow}$ state, the magnitude of the EE at the Fermi level is approximately $2.0\times{10}^{-9}\hbar \textup{m}/\textup{V}$. This value is comparable to that of 2D Rashba electron gases\cite{REE8,REE9} and 2D heterostructures\cite{l41n-qxpd}, yet one order of magnitude larger than those found in typical p-wave magnets such as LuFeO$_{3}$\cite{Manchon} and CeNiAsO\cite{Libor}. After normalizing by the longitudinal conductivity and considering the effective monolayer thickness of 4\AA, the EE coefficient reaches the magnitude of $1.62\times10^{11} \hbar/(\textup{A} \cdot \textup{cm})$ (see Appendix A for more details), which means that a charge current of 100 $\textup{A}/\textup{cm}^{2}$ will generate the spin density of 1.62$\times10^{12}/ \textup{cm}^{3}$ (or  6.48$\times10^{4}/ \textup{cm}^{2}$ after normalizing it into two-dimensional). With such a magnitude of EE, by adding a current of 10$^{6}$ to 10$^{8}$ $\textup{A}/\textup{cm}^{2}$ in monolayer PtBi$_{2}$, the spin polarization of the system could reach 0.1\% to 10\% (see Appendix B). 
Such a normalized EE coefficient is sizable compared to previously calculated chiral and ferroelectric systems, which typically fall within the range of $10^{9}\sim10^{10} \hbar/(\textup{A} \cdot \textup{cm})$\cite{FE3,FE4,FE6,REE17}.  It is also worth noting that the maximum value of the EE coefficient occurs at $E_{F}$=-0.05 eV, which lies near the Fermi level and can, in principle, be accessed by gating\cite{PhysRevLett.117.106801}. Meanwhile, when the Fermi level shifts toward higher energies, the EE for the P$_{\uparrow}$ (P$_{\downarrow}$) state decreases (increases) monotonically, and changes sign at $E_F=0.2$ eV.

Since the EE is dominated by the Fermi surface states\cite{PhysRevLett.128.166601}, it is essential to examine how the electronic states evolve with respect to the Fermi level, so that the microscopic origin of EE can be further unveiled. For this purpose, we evaluate the Fermi surfaces and corresponding spin textures at different chemical potentials for the P$_{\uparrow}$ state of monolayer PtBi$_{2}$, as shown in Fig. \ref{3}. As clearly seen in Fig. \ref{3}(b), at $E_F$=0 eV, the two electron pockets near the $\Gamma$ point exhibit Rashba-like spin textures with opposite helicity. In addition, six separate hole pockets appear away from the $\Gamma$ point. When the Fermi level is shifted downward to $E_{F}$ = -0.05 eV (see Fig. \ref{3}(a)), the spin textures of the inner electron pockets remain qualitatively unchanged, though their areas shrink. Meanwhile, the outer hole pockets expand and begin to merge. This not only makes the outer hole bands away from the $\Gamma$ point more pronounced, but also leads to the formation of a closed wrapped hole pocket, with a Rashba-like spin texture. This behavior arises because lowering the Fermi level allows additional hole bands to occur at the Fermi surface. In contrast, when the Fermi level is shifted upward to $E_{F}$ = 0.2 eV (see Fig. \ref{3}(c)), the hole pockets shrink and eventually disappear, while the electron pockets with Rashba-like spin textures expand. Meanwhile, the outer electron pocket also develops a slight warping distortion, which is compatible with the $C_{3v}$ symmetry of the system\cite{PhysRevB.107.094422,PhysRevB.85.075404}. The above analysis applies to the P$_{\uparrow}$ state of monolayer PtBi$_{2}$. For the P$_{\downarrow}$ state, while the shape of the Fermi surface at each evaluated energy remains unchanged, the directions of the spin textures are reversed, which is not shown here. 

 \begin{figure*}[ht]
\includegraphics[scale = 0.38 ]{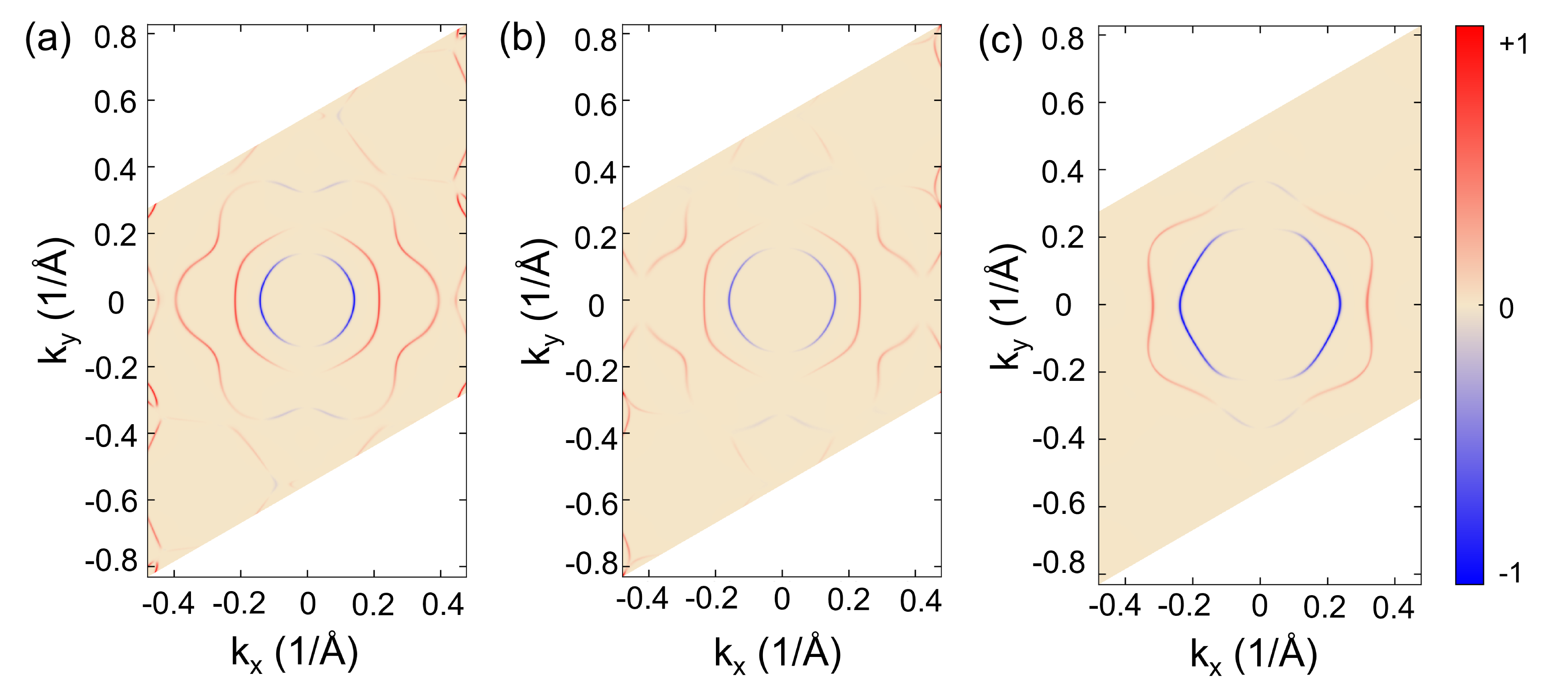}
\caption{\label{4} The $\mathbf{k}$-resolved EE coefficient ($\chi_{yx}(E_{F},\mathbf{k})$) for $P_{\uparrow}$ state of monolayer PtBi$_{2}$ at (a) $E_{F}$ = -0.05 eV, (b) $E_{F}$ = 0 eV, and (c) $E_{F}$ = 0.2 eV, respectively. The red and blue denote the positive and negative values of $\chi_{yx}(E_{F},\mathbf{k})$, respectively, which have been normalized to the maximum value.}
\end{figure*}

 \begin{figure*}[ht]
\includegraphics[scale = 0.32 ]{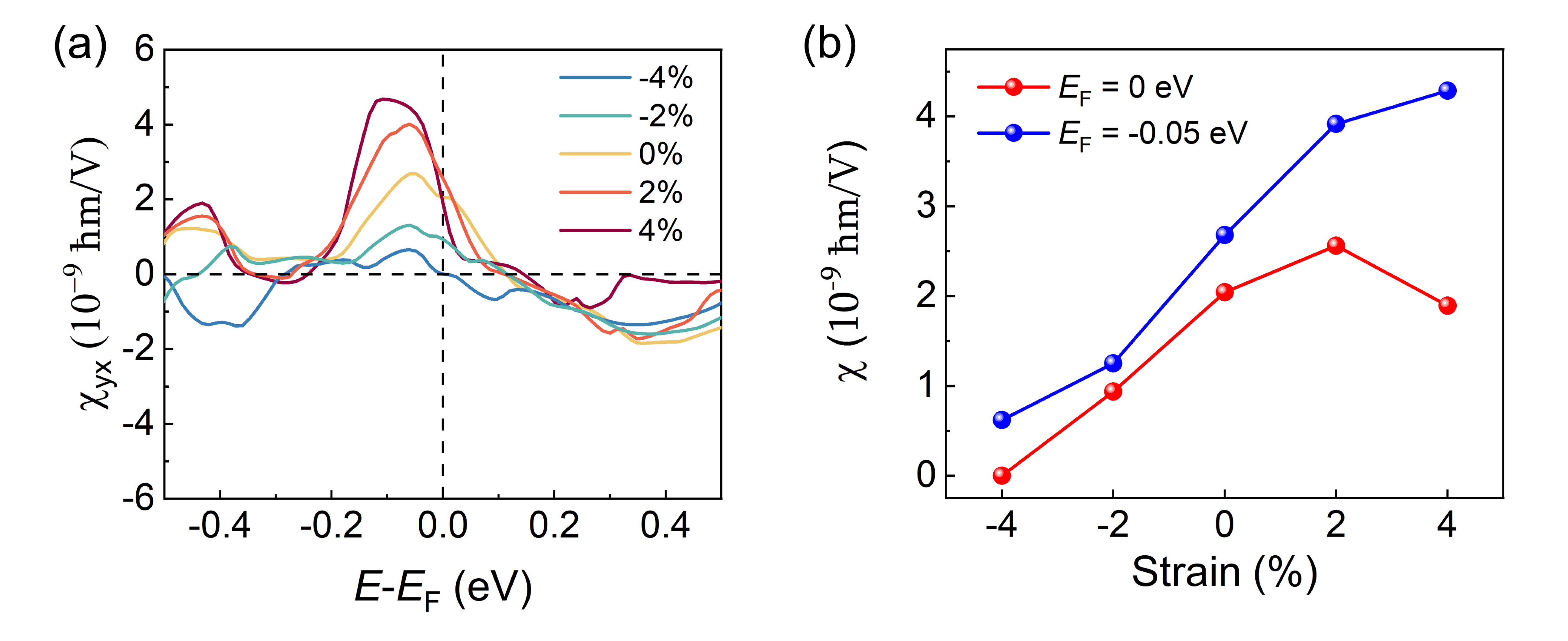}
\caption{\label{5} (a) The EE coefficient for $P_{\uparrow}$ state of monolayer PtBi$_{2}$ under different biaxial strain levels. (b) The value of EE coefficient at $E_{F}$ = 0 eV and $E_{F}$ = -0.05 eV as a function of biaxial strain level.  }
\end{figure*}

The above variation of the Fermi level modifies the distribution of electronic states on Fermi surface and thus alters their contribution to the EE. To gain deeper insights, we rewrite the EE into a sum of k-resolved terms:

\begin{equation}
\begin{aligned}
    \chi_{yx} = \frac{(2\pi)^{3}}{VN} \sum_{k} \chi_{yx}(E_{F},\mathbf{k})
\end{aligned}
\end{equation}

where $V$ and $N$ is the volume of the unit cell and the total number of k points used to sample the Brillouin zone, respectively. The distribution of k-resolved EE coefficient $\chi_{yx}(E_{F},\mathbf{k})$ for the  PtBi$_{2}$ monolayer in P$_{\uparrow}$ state is shown in Fig. \ref{4}. As seen at $E_F$=0 eV (see Fig. \ref{4}(b)), the two inner electron pockets contribute oppositely to the current-induced spin polarization, which is in line with their opposite Rashba-like spin textures. Meanwhile, the outer hole pockets generally contribute positively to the EE. Therefore, the total EE in monolayer PtBi$_{2}$ arises from the interplay between the inner electron pockets near the $\Gamma$ point and the outer hole pockets. When the Fermi level is shifted downward to $E_F$ = -0.05 eV  (see Fig. \ref{4}(a)), the contribution from the inner electron pockets remains qualitatively unchanged, whereas more hole states with positive contributions appear on the Fermi surface, thereby positively enhancing the total EE. Conversely, when the Fermi level is shifted upward, the outer hole pockets gradually shrink and eventually disappear at $E_{F}$=0.2 eV  (see Fig. \ref{4}(c)). In this situation, the EE is dominated by the two inner electron pockets, whose combined contribution is negative. As the outer hole pockets are responsible for the positive component of the EE, their disappearance suppresses the positive contribution, leading to a monotonic decrease of the total EE and ultimately causing its sign reversal as the Fermi level increases from 0 eV to 0.2 eV. Based on the above analysis, we conclude that the EE in monolayer PtBi$_{2}$ is affected by the delicate interplay between inner electron pockets and outer hole pockets. Moreover, by tuning the Fermi level, which can be experimentally achieved through slight doping or gating\cite{PhysRevLett.117.106801}, one can modulate the competition between contributions from the inner electron pockets and the outer hole pockets, thereby controlling the Edelstein effect in monolayer PtBi$_{2}$.

Besides ferroelectric switching, the EE in realistic ferroelectric systems can also be manipulated by applying moderate strain\cite{FE6}, which can be achieved by growing the sample on an appropriate substrate, where the lattice mismatch between the substrate and the sample induces and sustains strain\cite{strain,strain2,strain3}. This naturally raises the question of whether the EE in monolayer PtBi$_{2}$ is tunable by strain. To address this, we evaluate the EE coefficient in monolayer PtBi$_{2}$ as a function of biaxial strain, as shown in Fig. \ref{5}(a). It is evident that the EE coefficient is strongly affected by strain. For example, tensile strain generally enhances the EE across the entire evaluated energy range, whereas compressive strain suppresses it. By plotting the EE coefficient as a function of strain at $E_F$=0 eV, which lead to the intrinsic value of EE without the need of doping, and at $E_F$ =-0.05 eV, where the pristine monolayer exhibits its maximum EE, respectively (see Fig. \ref{5}(b)), one finds that strain induces a substantial variation in the EE. Specifically, for $E_F$ = 0 eV, the tensile strain does not lead to much change of the EE coefficient, while the compressive strain would strongly suppress the EE. To be specific, a compressive strain of -2\% will suppress the EE for about 50\%, while adding a larger compressive strain of -4\% will make the EE decrease to roughly 0. On the other hand, for $E_F$=-0.05 eV, a moderate tensile strain of 4\% enhances the EE to about 150\% of its unstrained value, while a compressive strain of −4\% reduces it to roughly 30\%. Microscopically, these strain-induced variations of EE arise from changes in the electronic states on the Fermi surface, which in turn modify their contributions to the EE (see Appendix C). These results highlight the potential of monolayer PtBi$_{2}$ as a platform for strain-tunable charge-to-spin conversion.

\section{Conclusions}

In conclusion, based on first-principles calculations, we have explored the EE in the intrinsic 2D ferroelectric metal PtBi$_{2}$. We find that the PtBi$_{2}$ monolayer exhibits sizable intrinsic EE coefficients, whose signs can be reversed through ferroelectric switching, thereby enabling non-volatile electrical control of charge–spin conversion. Microscopically, the EE originates from the interplay between the inner Rashba-like electron pockets and the outer hole pockets on the Fermi surface. Such interplay can be altered via shifting the Fermi level, thereby leading to a sign change of EE when the Fermi level is shifted upwards. Furthermore, by adding strain to the system, the electronic state on the Fermi surface would be strongly modified, thereby leading to a pronounced variation in EE. We believe that our results not only broaden the material candidates for effective charge–spin conversion in low-dimensional systems, but also pave the way toward non-volatile, electrically controllable spintronic devices.

\section{Acknowledgments}
This project was supported by the European Union Graphene Flagship project 2DSPIN-TECH (grant agreement No. 101135853) and SFB 1277 (Project-ID 314695032).

\section{Appendix A: Normalizing the EE coefficient}
The EE can be redefined as\cite{li2020out,REE17}:  
\begin{equation}
     \chi^{'}_{ij} = \delta \mathbf{s}_{i}/ \mathbf{J}_{j} =  \delta \mathbf{s}_{i}/ (\sigma_{jj} E_{j}   )
\end{equation}

where $\sigma_{ii}$ is the longitudinal conductivity of the system, which can be calculated using the Kubo formula implemented in Linres code\cite{linears}:

\begin{equation}
    \sigma_{ii} = \frac{e^{2}\hbar}{\pi V N} \sum_{\mathbf{k},m,n} \frac{\Gamma^{2} \textup{Re}(\langle n\mathbf{k} |\hat{v}_{i} | m\mathbf{k} \rangle \langle m\mathbf{k} | \hat{v}_{i} | n\mathbf{k} \rangle  )  }{[(E_{f}-\epsilon_{n\mathbf{k}})^{2}+ \Gamma^{2}][(E_{f}-\epsilon_{m\mathbf{k}})^{2}+ \Gamma^{2}]}
    \label{eq1}
\end{equation}

Here, $e$ is the elementary charge; $n$ and $m$ denote band indices; $\textbf{k}$ is the Bloch vector; $E_{F}$ is the Fermi energy; $\hat{v}$ is the velocity operator, $\hat{S}$ is the spin operator; $\epsilon_{n\textbf{k}}$ is the eigenvalue; $V$ is the volume of unit cell; $N$ is the total number of k points used to sample the Brillouin zone; and $\Gamma$ = 10 meV is the disorder parameter. Based on the calculated longitude conductivity and considering the monolayer thickness of about 4 \AA, we arrive at a renormalized EE of $\chi^{'}_{yx} = -\chi^{'}_{xy} = 1.62 \times 10^{11} \hbar/(\textup{A} \cdot \textup{cm})$. Note that the magnitude of the normalized Edelstein effect is largely independent of the $\Gamma$ broadening and can thus serve as a reliable reference value. 

\section{Appendix B: Estimating the current-induced spin polarization}
The spin polarization in a two-dimensional system is defined as $p = S/n$, where $S$ is the induced spin density under a specific current and can be obtained through the calculated EE coefficient, and $n$ is the carrier density. The $n$ can be estimated via Drude formula: $\sigma = n e^{2} \tau /m_{e}$, where $m_{e}$ is the mass of electron while $\tau = \hbar /2\Gamma$ is the relaxation time. Based on the longitudinal conductivity calculated with the Kubo formula in equation \ref{eq1}, and normalizing the calculated density into two dimensions, we can arrive at $n$ = $3.6\times 10^{11}$ cm$^{-2} $. With such an estimated $n$, we can predict the current-induced spin polarization for a given electrical current density stated in the main text. 

\section{Appendix C: The effect of strain on Fermi surface, spin textures, and EE effect}

\begin{figure*}
\includegraphics[scale = 0.4 ]{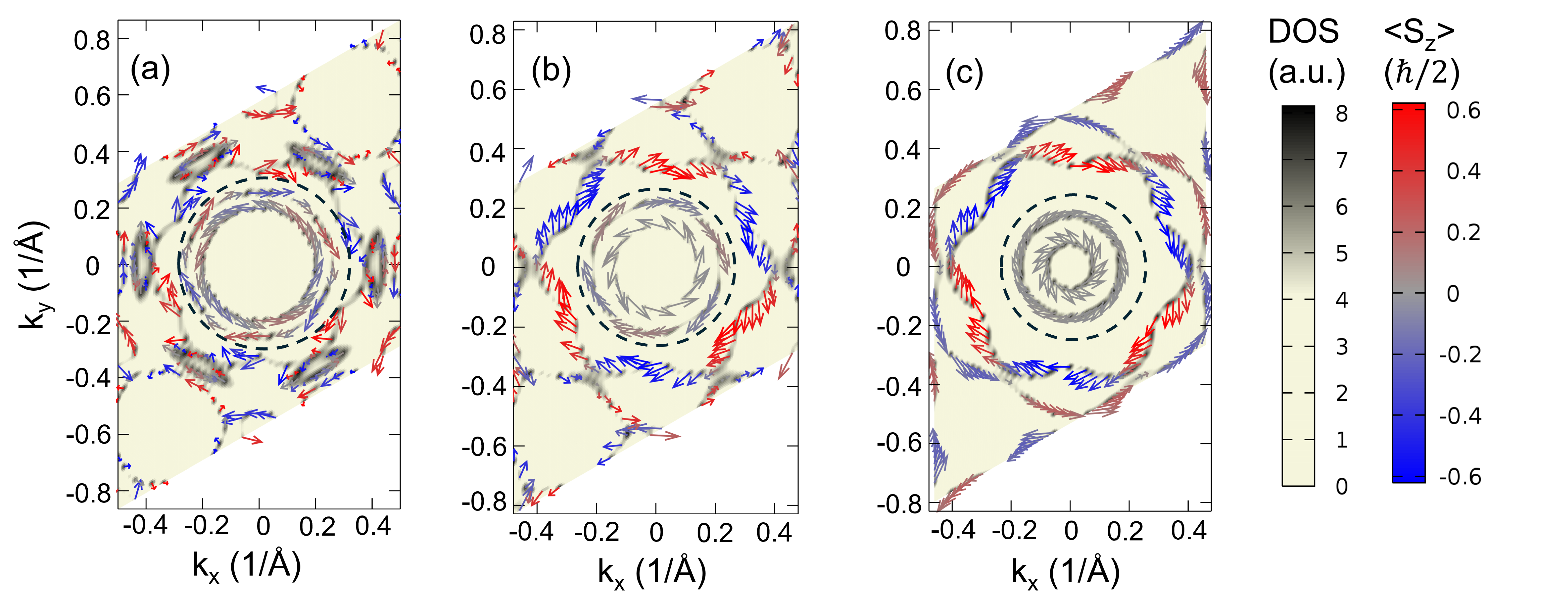}
\caption{\label{6} The Fermi surface states as well as the corresponding spin textures for P$_{\uparrow}$ state of monolayer PtBi$_{2}$ at $E_{F}$ = -0.05 eV with strain level of (a) -4\%, (b)  0\%, and (c)  4\%, respectively. The background color shows the spectral density of electronic states on the Fermi surface, expressed in arbitrary units. The arrows indicate the direction and magnitude of the in-plane spin expectation values for these Fermi surface states. The color of the arrow represents the expectation value of out-of-plane spin components. The states from inner electron pockets near the $\Gamma$ point are circled, while the states out of the circle are from outer hole pockets. }
\end{figure*}

\begin{figure*}[ht]
\includegraphics[scale = 0.4 ]{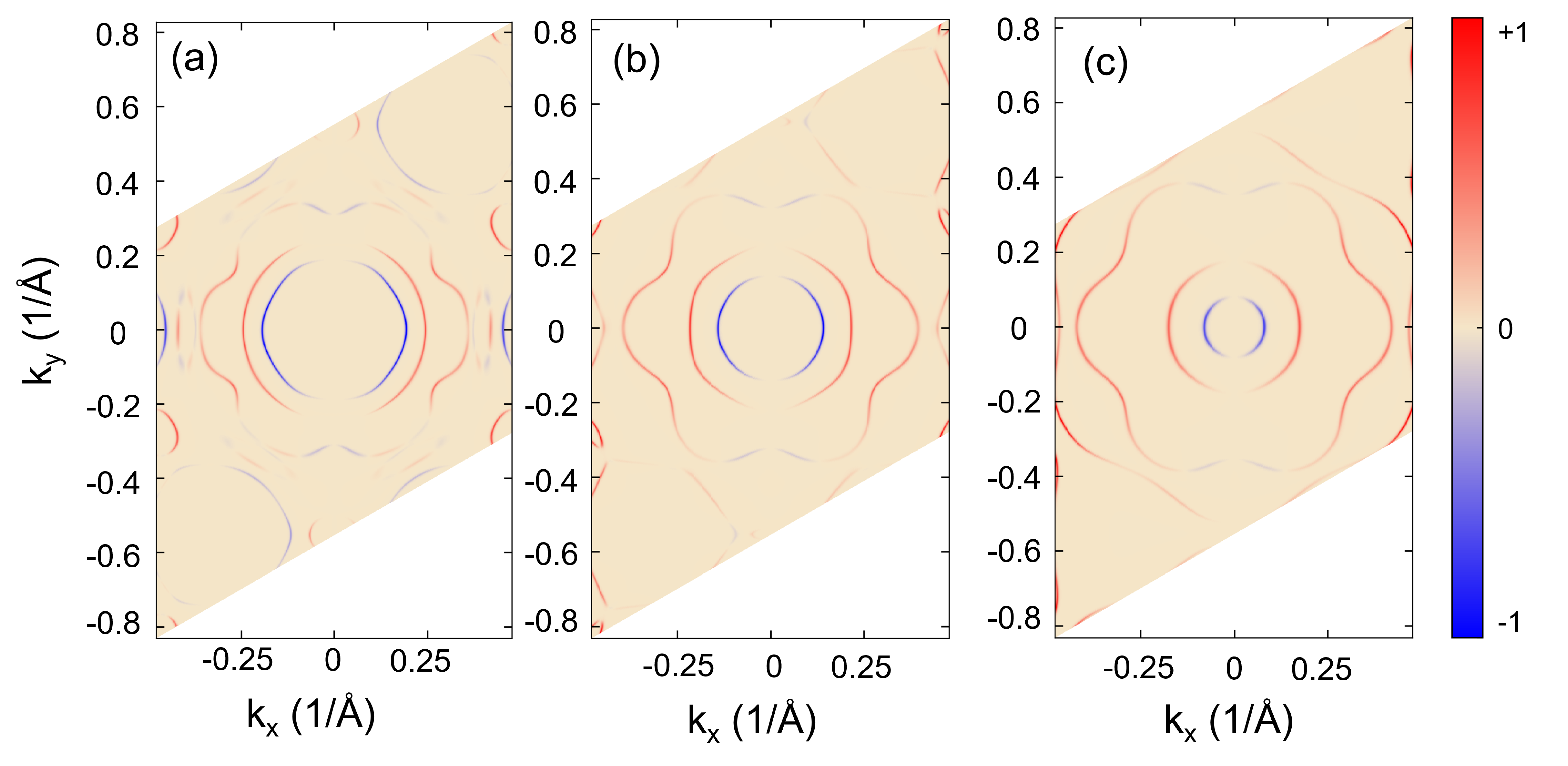}
\caption{\label{7} The $\mathbf{k}$-resolved EE coefficient ($\chi_{yx}(E_{F},\mathbf{k})$) for $P_{\uparrow}$ state of monolayer PtBi$_{2}$ at $E_{F}$ = -0.05 eV under strain level of (a) -4\%, (b) 0\%, and (c) 4\%, respectively. The red and blue denote the positive and negative values of $\chi_{yx}(E_{F},\mathbf{k})$, respectively, which have been normalized to the maximum value. }
\end{figure*}

\begin{figure*}[ht]
\includegraphics[scale = 0.35 ]{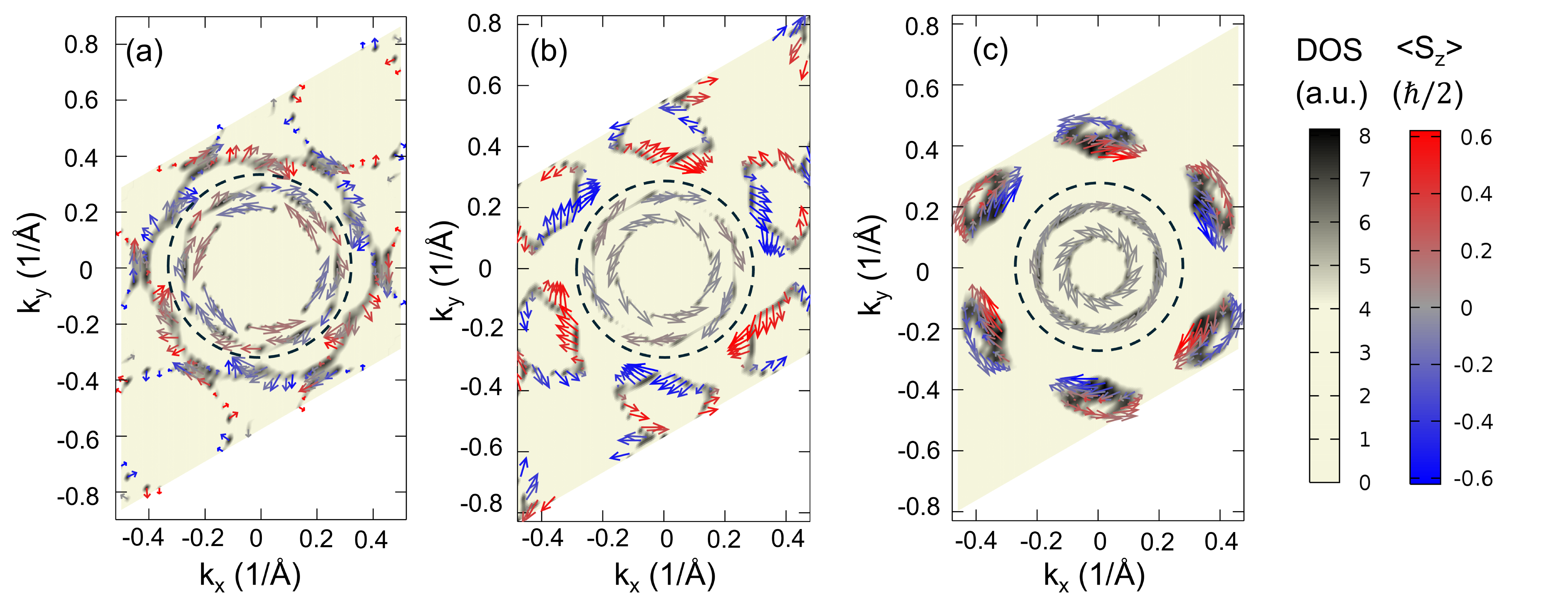}
\caption{\label{8} The Fermi surface states as well as the corresponding spin textures for P$_{\uparrow}$ state of monolayer PtBi$_{2}$ at $E_{F}$ = 0 eV with strain level of (a) -4\%, (b)  0\%, and (c)  4\%, respectively. The background color shows the spectral density of electronic states on the Fermi surface, expressed in arbitrary units. The arrows indicate the direction and magnitude of the in-plane spin expectation values for these Fermi surface states. The color of the arrow represents the expectation value of out-of-plane spin components. The states from inner electron pockets near the $\Gamma$ point are circled, while the states out of the circle are from outer hole pockets. }
\end{figure*}

\begin{figure*}[ht]
\includegraphics[scale = 0.37 ]{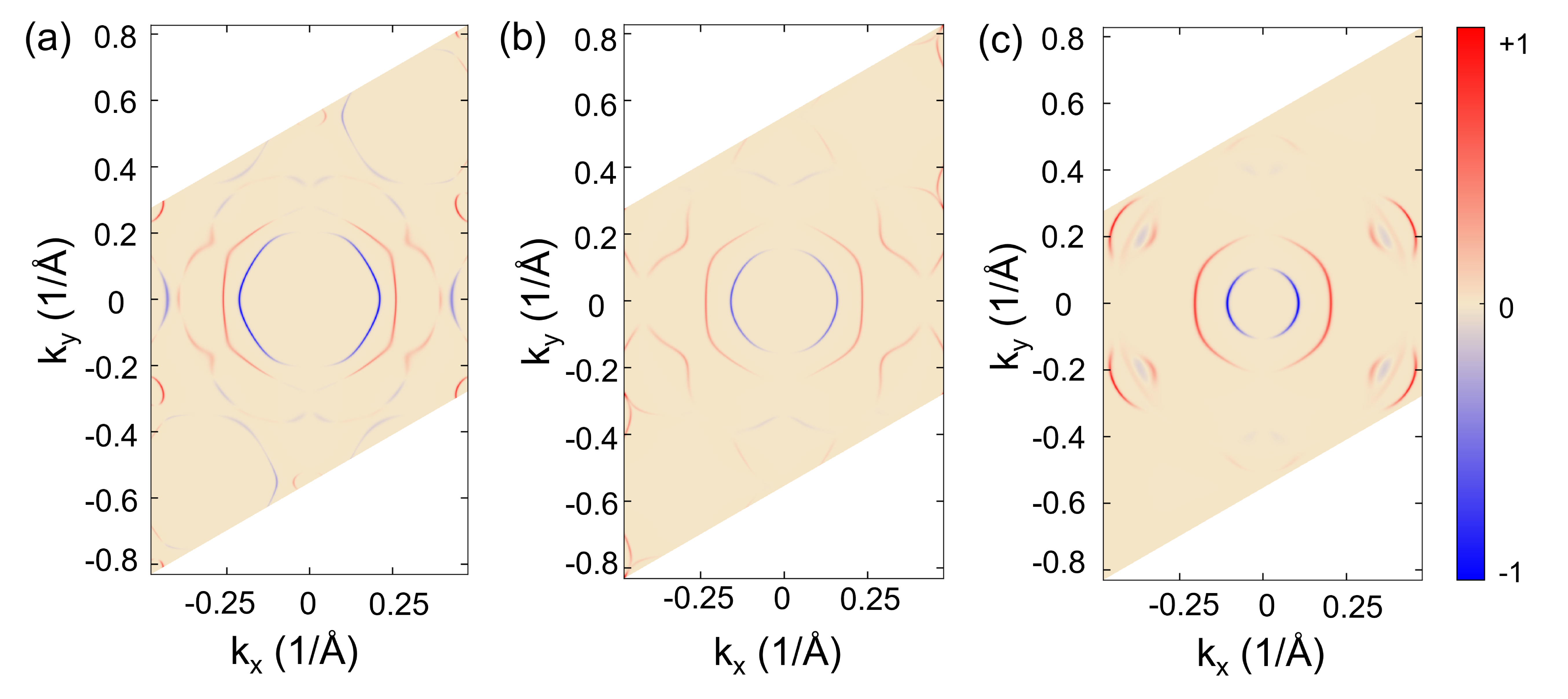}
\caption{\label{9} The $\mathbf{k}$-resolved EE coefficient ($\chi_{yx}(E_{F},\mathbf{k})$) for $P_{\uparrow}$ state of monolayer PtBi$_{2}$ at $E_{F}$ = 0 eV under strain level of (a) -4\%, (b) 0\%, and (c) 4\%, respectively. The red and blue denote the positive and negative values of $\chi_{yx}(E_{F},\mathbf{k})$, respectively, which have been normalized to the maximum value. }
\end{figure*}

Below, we present the spin textures of the Fermi surface states under strains of -4\%, 0\%, and 4\%  at 
$E_{F}$ = -0.05 eV, as shown in Fig. \ref{6}. It is evident that compressive (tensile) strain causes the inner electron pockets to expand (shrink), accompanied by the appearance of more (fewer) outer bands on the Fermi surface. Such strain-induced modifications of the Fermi surface significantly influence the 
k-resolved EE, which is shown in Fig. \ref{7}. A similar strain-dependent evolution is observed at 
$E_{F}$ = 0 eV (Fig. \ref{8}), where compressive (tensile) strain induces the expansion (shrinkage) of inner electron pockets and the emergence (reduction) of outer hole pockets, giving rise to the corresponding variation in the k-resolved EE distribution shown in Fig. \ref{9}.

\clearpage
%

\end{document}